\documentclass[final,authoryear,5p,twocolumn]{elsarticle}

\usepackage{subfig}

\usepackage{lineno}
\biboptions{longnamesfirst,comma}
\journal{Applied Radiation and Isotopes}

\begin{document}

\begin{frontmatter}

\title{Characterization of a $^{6}$Li-loaded liquid organic scintillator for fast neutron spectrometry and thermal neutron detection}

\author[NIST]{C.D. Bass\corref{JLab}}
%\ead{cbass@jlab.org}
\author[UMd]{E.J. Beise}
\author[UMd]{H. Breuer}
%\author[UMd]{D.K. Erwin}
\author[NIST]{C.R. Heimbach}
\author[UMd,IREAP]{T. Langford}
\author[NIST]{J.S. Nico}
%\ead{jeffrey.nico@nist.gov}
%\author[UMd]{A. Rodrigues}

\address[NIST]{National Institute of Standards and Technology, Gaithersburg, MD 20899, USA}
\address[UMd]{Department of Physics, University of Maryland, College Park, MD 20742, USA}
\address[IREAP]{Institute for Research in Electronics and Applied Physics, University of Maryland, College Park, MD 20742, USA}
\cortext[JLab]{Corresponding author. Present address: Thomas Jefferson National Accelerator Facility, Newport News, VA 23606, USA}

\begin{abstract}
The characterization of a liquid scintillator incorporating an aqueous solution of enriched lithium chloride to produce a scintillator with 0.40\% $^{6}$Li is presented, including the performance of the scintillator in terms of its optical properties and neutron response.  The scintillator was incorporated into a fast neutron spectrometer, and the light output spectra from 2.5 MeV, 14.1 MeV, and $^{252}$Cf neutrons were measured using capture-gated coincidence techniques.  The spectrometer was operated without coincidence to perform thermal neutron measurements.  Possible improvements in spectrometer performance are discussed.
\end{abstract}


\end{frontmatter}

\pdfoutput=1

\section{Introduction}

Precise knowledge of the fast neutron flux and spectrum is essential for several experimental endeavors requiring a low-background underground environment \citep{For04}.  These include searches for WIMP dark matter \citep{Ake03,Ang05,Gai04}, neutrinoless double beta decay \citep{Aal05,Ell02,Sch06}, and solar neutrinos \citep{Abd09,Aha07,Cle98,Fuk01,Ham99,McK05}.  The technological challenges associated with fast neutron spectrometry in underground labs are similar to those presented to fast neutron flux measurements for detecting and identifying fissile materials with low-level neutron activity.  Both applications require a detector with a low energy threshold, high sensitivity, and good energy resolution.

Liquid organic scintillators are often used for fast neutron spectrometry because of their fast response times, good pulse-height response, and modest cost.  However, organic scintillators have a high gamma-ray sensitivity with comparable detection probabilities for neutrons and gamma-rays.  For certain types of organic scintillators, pulse-shape discrimination techniques can be used to distinguish between particle types \citep{Sod08}.  Even so, complicated unfolding procedures \citep{Kle02} are needed for obtaining energy information from pulse-height spectra because only a fraction of the high-energy particles are brought to rest in the scintillator.

One technique for performing fast neutron spectrometry that overcomes spectral unfolding procedures involves a capture-gated coincidence between an incident fast neutron that is completely thermalized within an organic scintillator and its subsequent capture on a nucleus with a large neutron capture cross section, which is loaded within the scintillator volume \citep{Czi02,Dra86,Kim10}.  Neutron thermalization is fast (on the order of a few nanoseconds), but the capture time is typically tens to hundreds of microseconds and depends on the diffusion time needed for a thermalized neutron to propagate through the scintillator and capture on a nucleus.  A capture signal preceded by a thermalization signal within a characteristic time can be used to select those fast neutrons that have deposited all of their kinetic energy into the scintillator, and the initial thermalization signal provides energy information about the incident neutron.

Both $^{10}$B and $^{6}$Li have large cross sections for thermal neutron capture.  The $^{10}$B$(n,\alpha)^{7}$Li reaction has a \emph{Q}-value energy of 2.79 MeV and produces either a 1.78 MeV alpha (ground state, 6\% branching ratio) or a 1.47 MeV alpha and a 477 keV gamma-ray (first excited state, 94\% branching ratio).  The $^{6}$Li$(n,\alpha)^{3}$H reaction has a \emph{Q}-value energy of 4.78 MeV and produces a 2.05 MeV alpha and a 2.73 MeV triton.  Because the fluorescent efficiency in organic scintillators generally decreases for heavier particles \citep{Bir51}, a $^{6}$Li loading should produce a higher light output\footnote{It is convenient to express light output in terms of the scintillator's response to electrons, as this can be taken to be linear at least above about 100 keV \citep{Fly64}.} for neutron capture than a $^{10}$B loading.  For example, the calculated light output for the neutron capture products from $^{6}$Li and $^{10}$B in BC501A\footnote{Certain commercial equipment, instruments, or materials are identified in this paper in order to specify the experimental procedure adequately. Such identification is not intended to imply recommendation or endorsement by the National Institute of Standards and Technology, nor is it intended to imply that the materials or equipment identified are necessarily the best available for the purpose.} (a liquid organic scintillator based on xylene, which is commonly used in neutron spectroscopy and produced by Saint-Gobain) is shown in Table~\ref{light}.  The light output of the triton from $^{6}\textrm{Li}(n,\alpha)^{3}\textrm{H}$ is around an order of magnitude larger than that of the alpha from $^{10}\textrm{B}(n,\alpha)^{7}\textrm{Li}$.

While organic scintillators loaded with boron or gadolinium are available, to our knowledge there are currently no commercially-available sources of lithium-loaded organic scintillators.  There are other methods for adding $^{6}$Li to a scintillator (e.g. immersion of $^{6}$Li-glass plates in BC-501A \citep{Hay06}), but those techniques differ significantly in that the $^{6}$Li is not distributed uniformly throughout the volume, as is the case with most other commercially-available scintillators that are loaded with boron or gadolinium.  This is particularly important if one wants to scale up a detector, where optical clarity becomes important.  In this paper, we discuss the development, production, and optical characterization of a $^{6}$Li-loaded liquid organic scintillator along with measurements of its response to fast and thermal neutrons.

%\begin{table}[b]
\begin{table}
    \centering
    \begin{tabular}{cccc}
        & neutron & & Light \\
        & capture & Energy & Output \\
        & product & (MeV) & (keV$_{ee}$) \\
        \hline
        $^{10}$B$(n,\alpha)^{7}$Li & $\alpha$ & 1.78 & 56 \\
        $^{10}$B$(n,\alpha)^{7}$Li$^{*}$ & $\alpha$ & 1.47 & 37 \\
        & $\gamma$ & 0.477 & 311 \\
        $^{6}$Li$(n,\alpha)^{3}$H & $\alpha$ & 2.05 & 74 \\
        & \emph{t} & 2.73 & 409 \\
        \hline
    \end{tabular}
    \caption{Calculated light output for products of thermal neutron capture on $^{10}$B and $^{6}$Li in the organic liquid scintillator BC501A.  The light output of the charged particles was calculated using the Bethe-Bloch formula and measured data \citep{Ver68,Nak95}.}
    \label{light}
\end{table}

\section{Lithium-loaded liquid scintillator chemistry}

Loading an organic scintillator with $^{6}$Li can be accomplished by either dissolving a lithium compound directly into the scintillator solvent or by incorporating a medium containing a lithium compound into the scintillator bulk, where the medium does not dissolve into the solvent.  Aqueous solutions of lithium compounds are technically straightforward to produce although they will not dissolve into the aromatic solvents of organic scintillators.  However, there exists a class of organic liquid scintillators that was developed specifically for accepting aqueous solutions by the inclusion of surfactants \citep{Wie91}.  These so-called ``scintillator cocktails'' are predominantly used for biological and health physics applications and they possess good light output characteristics, low toxicity, a high flashpoint, and are economically priced.

Zinsser Analytic developed a high-efficiency scintillator cocktail under the commercial name Quickszint 164 that could incorporate aqueous solutions up to 40\% water by volume.  It is a mixture of an aromatic solvent (di-isopropyl naphthalene), organic fluors (PPO and bis-MSB), a non-ionic surfactant (ethoxylated nonylphenol), and mineral oil.  Quickszint 164 has a high flashpoint ($>$100$^\circ$ C), is nontoxic, and is biodegradable.  It has a density of 0.92 g cm$^{-1}$, viscosity of 27.6 cP at 20$^\circ$ C, index of refraction of 1.57, and a hydrogen-to-carbon ratio of 1.48 \citep{Zin}.

It should be noted that Quickzint 164 was specifically developed by Zinsser Analytic as an experimental scintillator cocktail under the project name XLS164H and is a non-stock item.  Zinsser Analytic currently produces Quickszint Flow 302+, which is a commercially-available liquid scintillator cocktail that is formulated using the same aromatic solvent, organic fluors, and non-ionic surfactant as Quickszint 164 with similar proportional composition.  We have not evaluated Quickszint Flow 302+ as a direct replacement for Quickszint 164.

\begin{table}[t]
%\begin{table}[htbp]
    \centering
    \begin{tabular}{lcc}
        & & maximum \\
        & solubility & Molarity \\
        & (mass \% of solute ) & (M) \\
        \hline
        Li$_{2}$CO$_{3}$ & 1.3 & 0.2 \\
        LiCl & 45.8 & 13.9 \\
        LiBr & 65.4 & 17.6 \\
        \hline
    \end{tabular}
    \caption{Solubility of lithium compounds in water at 25$^\circ$ C. Data taken from CRC Handbook \citep{CRC}.}
    \label{solubility}
\end{table}

Lithium chloride was chosen to incorporate into Quickszint 164 because of its solubility and maximum molar concentration in water \citep{Fis11} (see Table~\ref{solubility}).  For purposes of initially characterizing the optical and physical properties of a loaded scintillator, unenriched lithium compounds were used (isotope abundance for naturally-occurring lithium: 7.7\% $^{6}$Li, 92.3 \% $^{7}$Li).  Lithium carbonate was reacted with hydrochloric acid to produce aqueous lithium chloride.  Excess water was removed from the solution by a combination of heating and low vacuum, and the resulting slurry was diluted with deionized water to produce three batches of aqueous lithium chloride in different molar concentrations: 2.26 M, 4.72 M, and 9.48 M.  Each of these solutions was added to Quickszint 164 in varying volume fractions to create samples of lithium-loaded scintillator over a range of loadings.  The prepared samples were mixed on a roller-mixer for 10 minutes and then allowed to settle for 24 hours to allow trapped air bubbles to escape.

When aqueous solutions are added to scintillator cocktails, they form water-in-oil microemulsions that are uniform mixtures of oil, water, and surfactant, which occur with minimal mixing and are highly stable \citep{Fan09}.  Microemulsions are dynamical systems in which droplets undergo collision, fusion, and core material exchange.  However, the exchange process is controlled by activation energy and not diffusion, so an equilibrium droplet size and shape is maintained.  Typical microemulsions have structural dimensions between 2 nm and 50 nm and are transparent in the near-UV and visible light range.

The amount of lithium chloride solution that Quickszint 164 was able to accept was initially determined by visual inspection of the prepared samples. Complete emulsification resulted in a water-clear, colorless liquid with a slight bluish fluorescent tinge.  Observation of cloudiness, films, or suspended slurry was evidence of phase separation at the macroscopic level or incomplete emulsification.  Based on these criteria, Quickszint 164 had a definite lower and upper limit of the volume fraction of solution that could be emulsified.  Prepared samples were monitored for long-term stability, and all samples that were completely emulsified remained stable after one year.

\begin{figure}
    \centering
    \includegraphics[width=\linewidth]{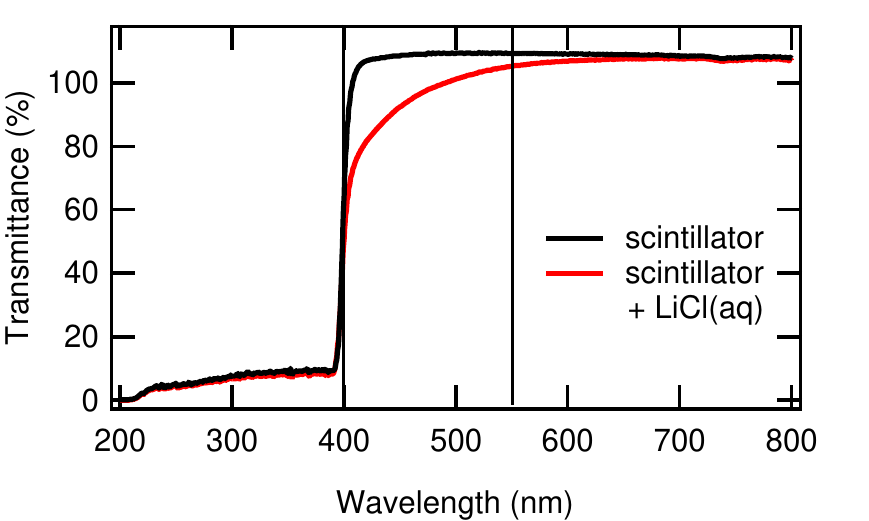}
    \caption{Transmittance spectra of Quickszint 164 (black line) and the 0.40\% $^{6}$Li-loaded scintillator (red line).  Vertical lines at 400 nm and 550 nm denote the relevant spectral band for fluorescence. Normalization was accomplished by taking a null baseline transmittance measurement of the empty sample cuvette.  Normalized transmittance values over 100\% are possible
    due to differences in refractive indices between the (quartz) sample cuvette, scintillator, and air.  The ratio of the transmittance integrals over the spectral band was 88\%, which indicated acceptable optical quality for the $^{6}$Li-loaded scintillator.}
    \label{UV-Vis}
\end{figure}

Lithium bromide was also investigated as a possible choice for an aqueous lithium compound because it has a larger maximum solubility in water and can form a higher-concentration solution than lithium chloride, which would allow a larger lithium-loading within Quickszint 164. A 12.3 M lithium bromide solution was prepared and added to Quickszint 164 in varying concentrations using the same procedures as the lithium chloride solutions.  However, when incorporated into Quickszint 164 over a range of lithium mass fractions equivalent to the upper range of the water-clear emulsions formed with lithium chloride, the lithium bromide solution formed a viscous, milky liquid that failed to emulsify.  In terms of emulsification, aqueous lithium chloride appears to be superior to aqueous lithium bromide for loading Quickszint 164, despite the higher solubility and larger maximum molar concentration of lithium bromide.

\section{Optical properties of the lithium-loaded scintillator}

The optical quality of each sample was evaluated by measuring its light transmittance across the UV-Vis spectrum.  The light transmittance \emph{T} for a sample is defined as

\begin{equation}
T(\lambda) = \frac{I(\lambda)}{I_{o}(\lambda)},
\end{equation}

\noindent where \emph{I} and $I_{o}$ are (respectively) the intensities of light transmitted through and incident upon a sample for a given wavelength $\lambda$. Transmittance measurements were performed with a spectrophotometer using a fused quartz, 10-mm pathlength cuvette.

Quickszint 164 uses bis-MSB as a waveshifter, which has an absorbance spectrum that ranges from near-UV to almost 400 nm and a fluorescence spectrum that ranges between 370 nm and 550 nm with prominent peaks at 400 nm and 424 nm.  The relevant spectral band for optical quality was therefore selected from 400 nm to 550 nm.

\begin{figure*}
   \centering
   \includegraphics[width=\linewidth]{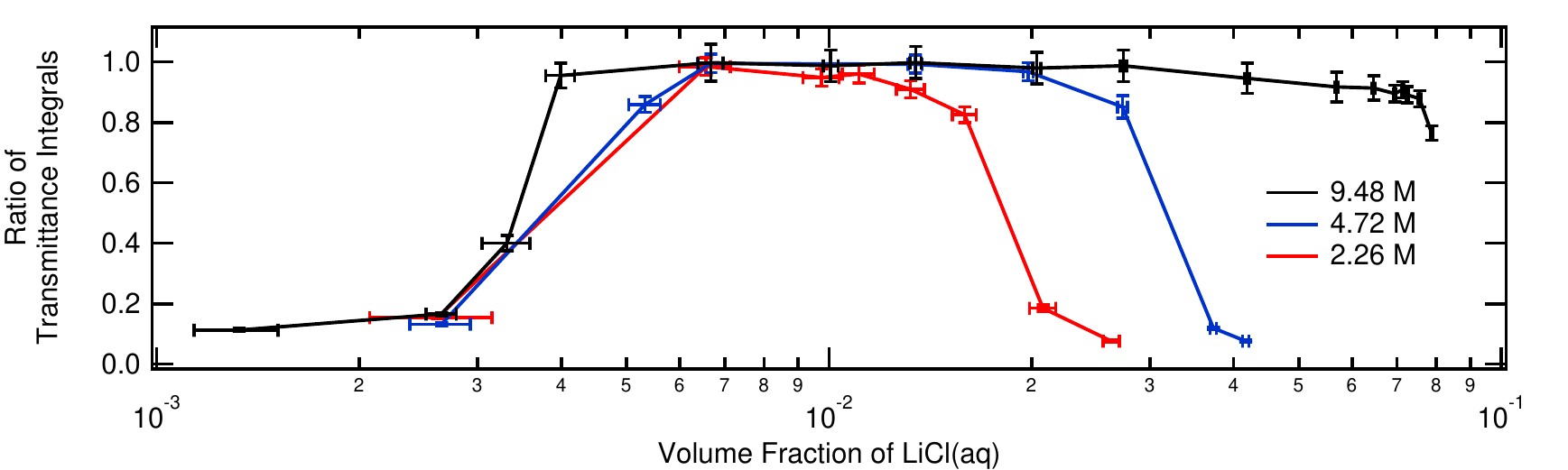}
   \caption{Plot of the optical performance of the Li-loaded scintillator as a function of the volume fraction of aqueous lithium chloride for various molar concentrations.  The minimal loading for acceptable transmittance is 0.4\% - 0.6\% aqueous lithium chloride by volume and is independent of molar concentration, while the maximal loading depends proportionally on the molar concentration of aqueous lithium chloride.  Error bars indicate the accuracy of the UV-Vis transmittance measurements and physical measurements during chemistry procedures.}
   \label{UV-Vis_analysis}
\end{figure*}

Transmittance measurements were taken for each sample, and a typical spectrum is shown in Figure~\ref{UV-Vis}.  The transmittance integral in the spectral band for each sample was compared to that of Quickszint 164, and the ratio of the integrals

\begin{equation}
\frac{\int{T_{Li}(\lambda)\, d\lambda} }{ \int{T_{pure}(\lambda)\, d\lambda}},
\end{equation}

\noindent provided a metric of the optical degradation of Quickszint 164 due to the inclusion of aqueous lithium chloride.

Light transmittance studies indicate a range of loadings in Quickszint 164 that display minimal optical degradation.  As shown in Figure~\ref{UV-Vis_analysis}, the lower loading limit depends primarily on the volume fraction of aqueous lithium chloride present in Quickszint 164. In contrast, the upper loading limit depends proportionally on the molar concentration of the aqueous lithium chloride -- highly-concentrated solutions extend the range of allowed loadings.

The choice of loading within the acceptable transmittance range is constrained by the neutron response of the loaded scintillator.  Monte Carlo simulations using MCNP5 \citep{Bro03} of a small volume detector employing $^{6}$Li-loaded scintillator indicate that detector efficiency increases with the mass fraction of $^{6}$Li while neutron diffusion time decreases.  Therefore, the optimal $^{6}$Li-loaded scintillator should incorporate enriched aqueous lithium chloride with a high molar concentration using the largest possible volume fraction in a scintillator cocktail, which possesses acceptable optical transmittance.

\section{Production of $^{6}$Li-loaded liquid scintillator}
\label{creation_of_scintillator}

A 9.40 M lithium chloride solution was prepared using lithium carbonate with a 95.5\% enrichment of $^{6}$Li.  It was mixed into Quickszint 164 to yield a lithium-loading of 0.40\% $^{6}$Li by mass and a calculated hydrogen-to-carbon ratio of 1.57.  UV-Vis measurements indicated a transmittance integral ratio of 88\%.

\begin{figure}
    \centering
    \includegraphics[width=\linewidth]{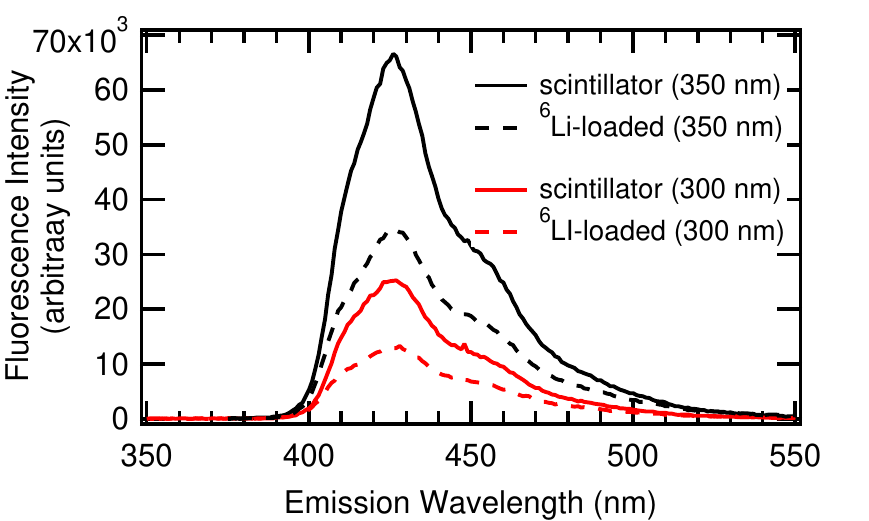}
    \caption{Fluorescence spectra for pure scintillator cocktail and $^{6}$Li-loaded scintillator using excitation energies of 300 nm and 350 nm.  The addition of aqueous lithium chloride decreases the fluorescence intensity of Quickszint 164 by about a factor of two.}
    \label{fluorescence}
\end{figure}

Because the addition of aqueous lithium chloride into Quickszint 164 reduced its light output, fluorescence measurements were performed to ascertain the level of decrease in scintillation light.  Fluorescence spectra were measured with a multifrequency phase fluorometer using single wavelength excitations at 300 nm and at 350 nm, which matched the fluorescence spectrum of the primary fluor, PPO.  As seen in Figure~\ref{fluorescence}, the addition of aqueous lithium chloride decreased the light output of Quickszint 164 by a factor of two.

The microemulsion droplet size of the $^{6}$Li-loaded scintillator was measured using a dynamic light scattering instrument.  The measured hydrodynamic droplet mean radius was 4.5 nm, which included the extent of the surfactant molecules from the droplet into the scintillator bulk.  The calculated mean length of the hydrophobic tail of the ethoxylated nonylphenol surfactant is approximately 1 nm, so the calculated diameter for a droplet of lithium chloride solution in Quickszint 164 is approximately 8 nm.

\begin{figure}
    \centering
    \includegraphics[width=\linewidth]{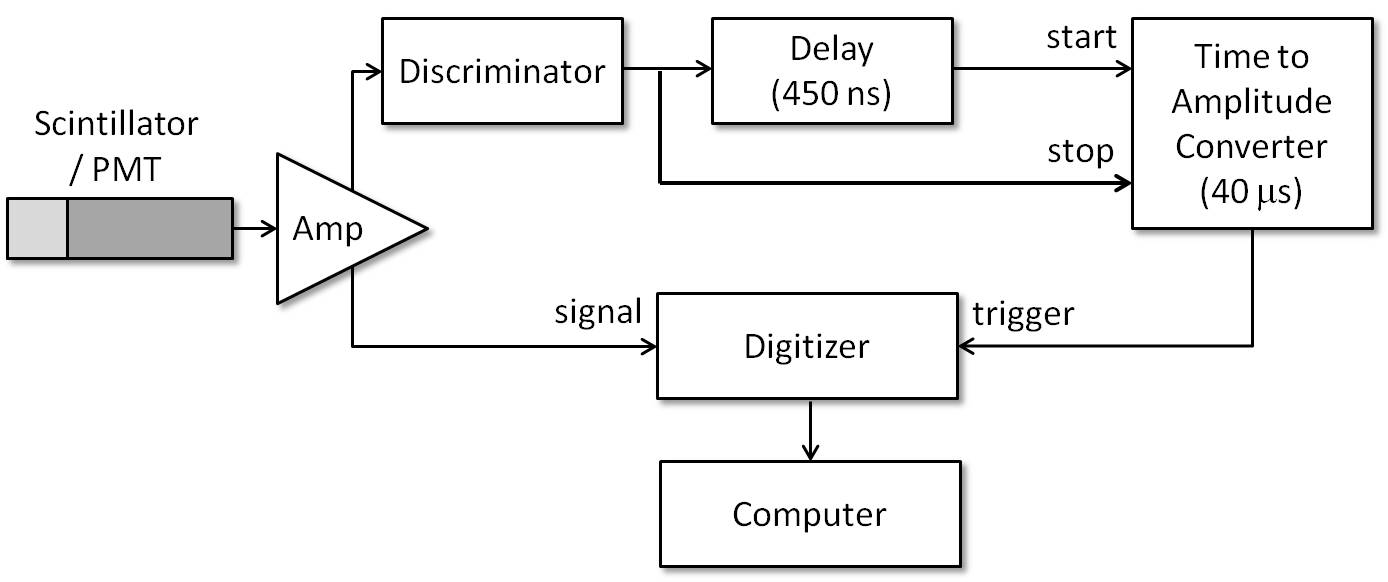}
    \caption{Block diagram of the data acquisition electronics. In capture-gated coincidence mode, the digitizer card is triggered by the time-to-amplitude converter (TAC) whenever a pair of PMT signals above threshold (set by a discriminator) are produced within 40 $\mu$s of each other.  The TAC is used as a coincidence analyzer with a longer resolving time than is possible with typical coincidence analyzers.  A 450 ns delay ensures that the coincidence trigger is due to a pair of pulses and not from the long tail of a single high-energy PMT signal.  In singles-mode, the digitizer card is triggered by any PMT signal above threshold (set by the card).  All digitized waveforms are recorded to disk for offline analysis.}
    \label{electronics}
\end{figure}

A test neutron detector was assembled by filling a 5 cm diameter by 6 cm cylindrical borosilicate glass cell (approximately 100 ml) with the prepared $^{6}$Li-loaded scintillator.  The volume and geometry of the glass cell was chosen 1) to couple with available 5 cm PMTs with characteristics deemed compatible with the expected performance of the $^{6}$Li-loaded scintillator, and 2) to provide roughly the same dimensions in terms of diameter and length so that the detector performance would be largely independent of neutron-field direction and easier to model in simulation.  The cell was externally coated with Bicron 622A reflective paint and coupled to a 5-cm Burle 8850 photomultiplier tube (PMT) using optical grease. PMT signals were recorded as waveforms by a GaGe 8-channel 125 MHz digital oscilloscope card with 14-bit resolution. The data acquisition electronics (see Figure~\ref{electronics}) allowed events to be recorded either in singles mode, where the digitizer triggered on signals above a threshold, or in capture-gated coincidence mode, where the digitizer would be triggered on any event (a ``start event'') above a threshold that was followed within 40 $\mu$s by a second event (a ``stop event'') above the threshold.  All of the digitized waveforms were recorded to disk for later analysis.

\section{Fast neutron response}

The PMT and glass cell containing the $^{6}$Li-loaded scintillator were inserted into a 0.64 cm thick lead cylinder to reduce the gamma-ray background.  This assembly was surrounded by 6 mm of borated-silicone to reduce the thermal neutron background.  Borated-silicone is a neutron shielding material based on a silicone elastomer that has boron carbide powder homogeneously mixed throughout its matrix.  The boron content attenuates thermal neutron flux due to its high thermal neutron absorption cross section.

Fast neutron irradiation of the detector was performed in the NIST Californium Neutron Irradiation Facility \citep{Gru77}, and neutrons were produced either by P-325 Neutron Generators manufactured by Thermo Electron Corporation (2.5 MeV neutrons from DD-fusion or 14.1 MeV neutrons from DT-fusion) or by spontaneous fission of $^{252}$Cf sources.  During irradiation, the detector was positioned approximately 2 m from a neutron source, and the resulting fast neutron field was a combination of an isotropic distribution from the source and neutron return from the environment, including albedo from boundaries of the irradiation room.  Data for each irradiation was acquired in capture-gated coincidence mode, which allowed the detector to function as a fast neutron spectrometer.

\begin{figure}
    \centering
    \includegraphics[width=\linewidth]{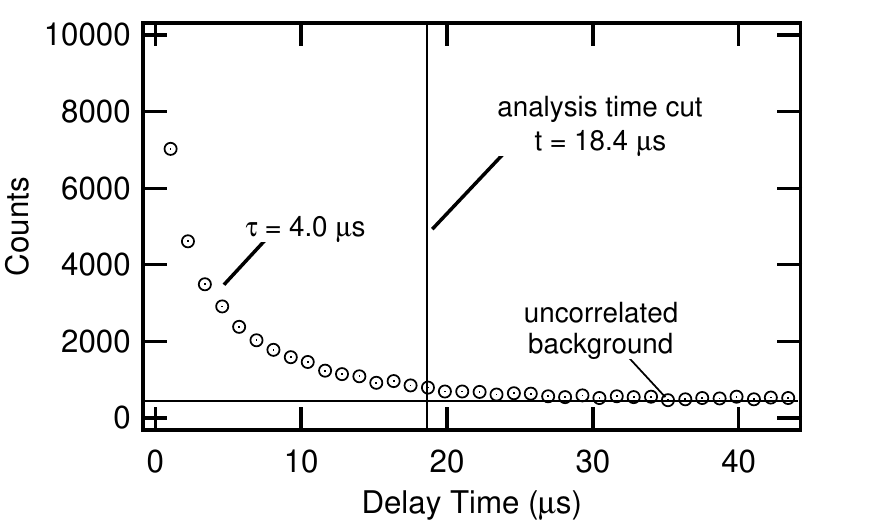}
    \includegraphics[width=\linewidth]{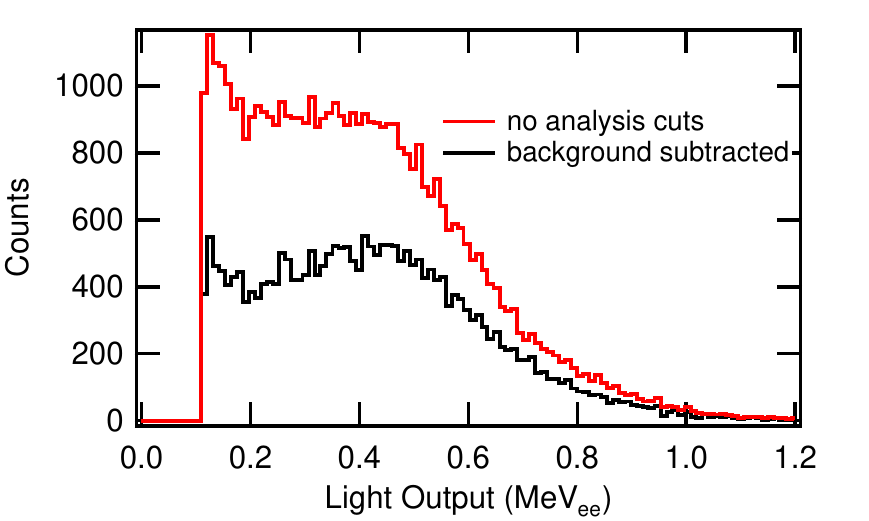}
    \includegraphics[width=\linewidth]{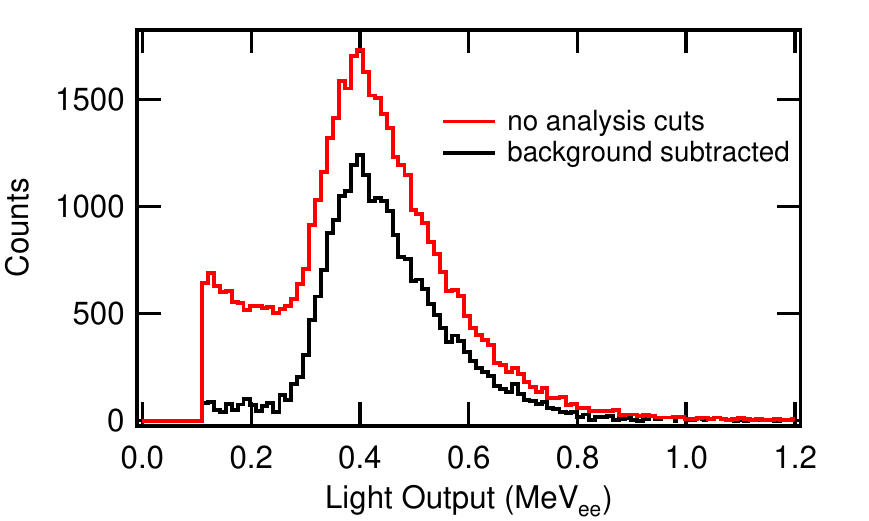}
    \caption{Background subtraction scheme used in offline analysis.  The top graph shows a typical distribution of time delays between the start and stop signals for capture-gated coincidence measurement of fast neutrons. This distribution is for an irradiation by 2.5 MeV neutrons and has a decay constant of 4.0 ${\mu}$s, which corresponds to the mean diffusion time in the scintillator for a thermalized neutron to capture on a $^{6}$Li nucleus.  The horizontal line indicates the level of uncorrelated background events due to random coincidences, and the vertical line corresponds to a time $t=4.6 \tau$, which is used to separate data from the background.  The middle and bottom graph shows the recoil and capture spectrum (respectively) for 2.5 MeV neutrons before and after the background subtraction cut in analysis.}
    \label{capture_times}
\end{figure}

Data runs of 50,000 events for each of the neutron sources were taken. Off-line analysis of the runs determined the relative time delays between start and stop signals for each digitized waveform as well as the pulse heights for the start and stop signals.  Histograms of the time delays for each run were created and fit with an exponential curve.  The decay constant $\tau$ for the fit curve corresponds to the mean diffusion time for a thermalized neutron to propagate through the scintillator and capture on a $^{6}$Li nucleus.  For this detector geometry and scintillator loading, the measured decay constant was 4.0 $\mu$s. Because coincidences with delay times greater than several mean diffusion times can be rejected as probable background events, a cut was applied to those coincidences that had delay times larger than $t=\tau \ln{100} \approx 4.6 \tau$, which should retain 99\% of non-background events (Figure~\ref{capture_times} shows a typical distribution).  Start and stop pulse height histograms of those cut events were used to perform a weighted background subtraction on the start and stop pulse height histograms for the remaining events.  This background subtraction scheme is used to improve both the signal-to-noise on the capture signal and energy resolution of the spectrometer.  Because the data were acquired in event mode, additional cuts on stop events that did not correspond to a neutron capture on $^{6}$Li could be made in analysis.

The detector was calibrated by removing the lead and borated-silicone and acquiring data in singles mode of irradiations by $^{60}$Co, $^{137}$Cs, $^{22}$Na, and $^{133}$Ba gamma-ray sources.  Pulse height information from each waveform was used to build histograms for each source, and the Compton edge of the gamma-rays for each source provided light output calibrations for the detector across a range of energies. The calibrations were then used to convert the start and stop event histograms into light output spectra for the proton recoil and neutron capture events (respectively).

\begin{figure*}
    \centering
    2.5 MeV Neutrons
    \subfloat{\includegraphics[width=0.5\linewidth]{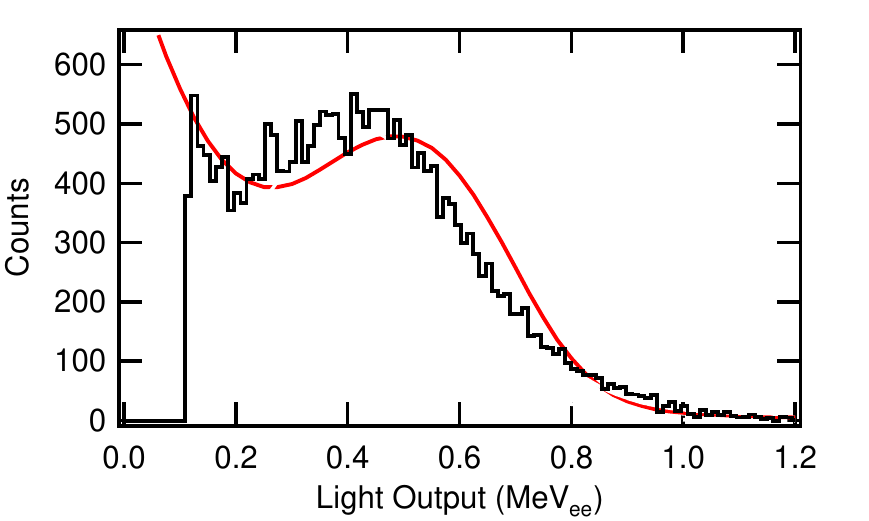}}
    \subfloat{\includegraphics[width=0.5\linewidth]{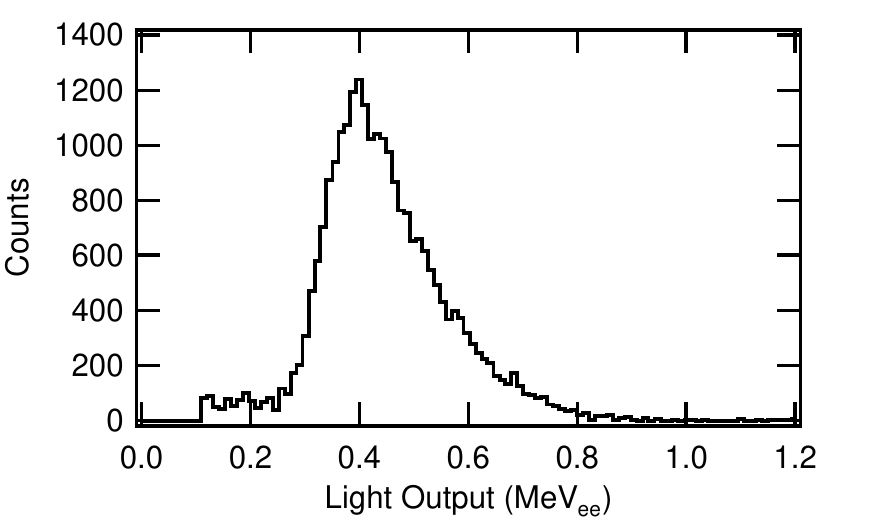}}\\
    14.1 MeV Neutrons
    \subfloat{\includegraphics[width=0.5\linewidth]{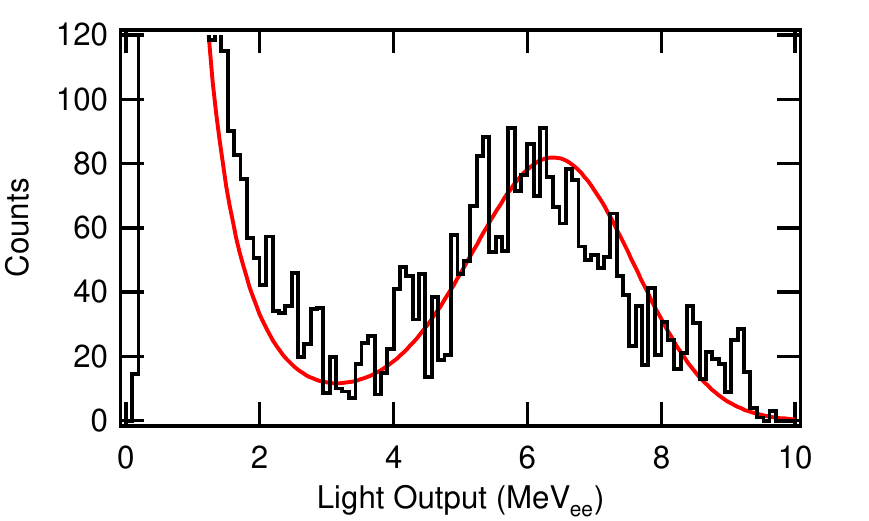}}
    \subfloat{\includegraphics[width=0.5\linewidth]{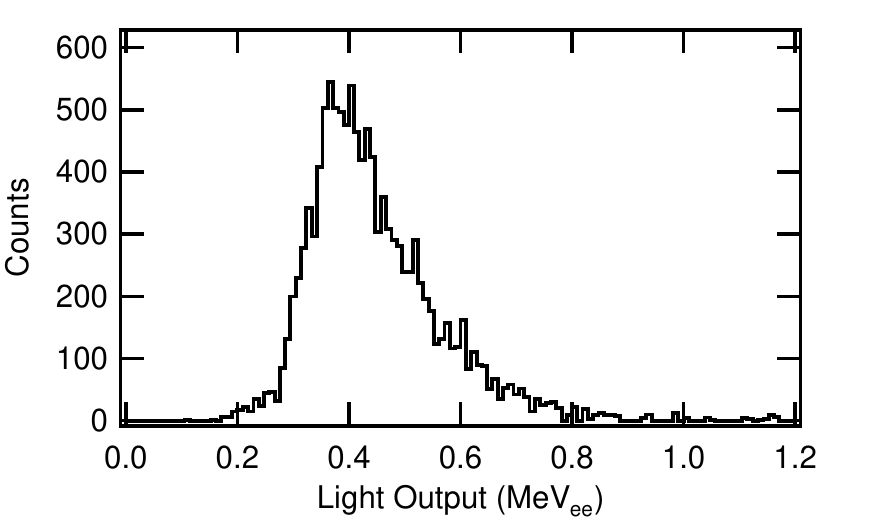}}\\
    Neutrons from $^{252}$Cf Decay
    \subfloat{\includegraphics[width=0.5\linewidth]{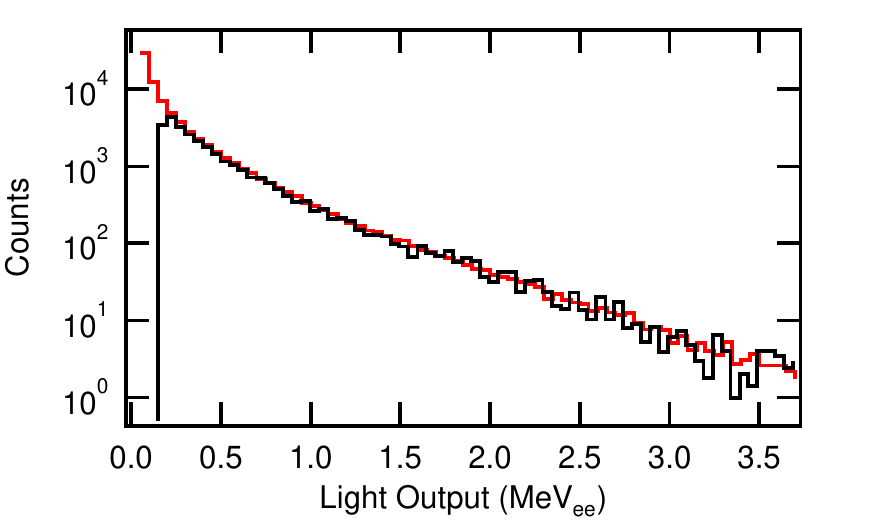}}
    \subfloat{\includegraphics[width=0.5\linewidth]{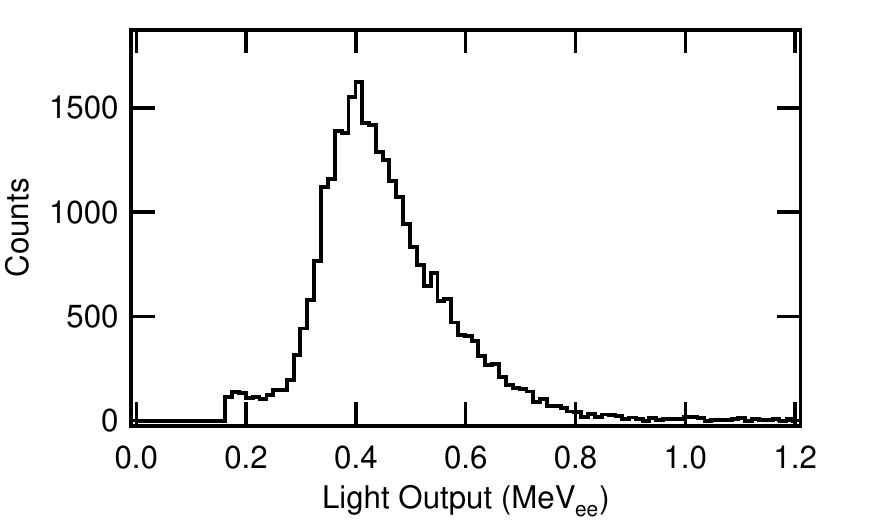}}
    \caption{Background-subtracted data from the $^{6}$Li-loaded scintillator test detector irradiated with 2.5 MeV neutrons (top graphs), 14.1 MeV neutrons (middle graphs), and neutrons from $^{252}$Cf decay (bottom graphs).  The left graphs shows the proton recoil light output spectra (black lines) that correspond to the energies of the incident neutrons and the Gaussian-smoothed Monte Carlo calculations (red lines) of the light output spectra (note: the low-energy portion of recoil spectra for the 14.1 MeV data and Monte Carlo were truncated because they reduce the clarity for seeing the peak around 6 MeV$_{ee}$, and because the exponential shape of data and Monte Carlo are unremarkable). The right graphs shows the neutron capture light output spectra, which corresponds to the total energy of the $^{6}$Li(n,$\alpha$)$^{3}$H reaction products. Data was collected at rates ranging between 0.4 events/s and 20 events/s. For clarity, the bin widths for the recoil spectra for 14.1 MeV neutrons and neutrons from $^{252}$Cf decay were increased by a factor of 10 and 4 (respectively) relative to their corresponding neutron capture light output spectra.}
    \label{fast_spectra}
\end{figure*}

The measured light output spectra for 2.5 MeV neutrons, 14.1 MeV neutrons, and neutrons from $^{252}$Cf decay are shown in Figure~\ref{fast_spectra}, which shows the proton recoil and neutron capture spectra for each neutron source after performing background subtraction in analysis.  Monte Carlo calculations of the light output spectra in Figure~\ref{fast_spectra} were performed using MCNP5 for the response functions and MCNP5's PTRAC option for identifying energy deposition with particle type.  The energy-to-light response functions from \citet{Ver68} for various charged particles were used to convert deposited energy to light on an event-by-event basis.

Neutron capture on $^{6}$Li verifies that all of the neutron interactions occurred within the scintillator and would indicate that the response should be a peak in the recoil spectra for each monoenergetic source and an approximately Maxwellian recoil spectrum for the $^{252}$Cf source.  Monte Carlo calculations of the recoil spectra included the free-field for each neutron source and the geometry and composition of the irradiation facility, and show a low-energy tail that primarily arises due to the neutron return in addition to the peaks for the monoenergetic sources and the Maxwellian distribution for the $^{252}$Cf sources.  This low-energy tail comes close to merging with the peak resulting from the 2.5 MeV neutrons but is well-separated from the 14.1 MeV peak; the low-energy tail merges completely with the Maxwellian distribution of the recoil spectrum for the $^{252}$Cf sources.  The calculations were smoothed using a Gaussian convolution and then overlaid on the measured spectra; the calculations show good agreement to the measured recoil spectra.

It should be noted that the measured data are of the light production in the scintillator and not of energy deposition.  The non-linear light yield in an organic scintillator \citep{Bir51} for hydrogen (protons) cause an apparent loss of energy, and the light yield for carbon recoils is significantly less than for proton recoils, which causes an additional loss of response that further contributes to the low-energy tail in the recoil spectra.  At higher energies, neutron inelastic reactions generate 4.4 MeV gamma-rays that are lost to the scintillator, which contribute to a gap between the peak and the low-energy tail in the 14.1 MeV neutron recoil spectrum.

\section{Thermal neutron studies}

Calculations done in MCNP5 indicate that the mean free path of thermal neutrons in the $^{6}$Li-loaded scintillator is 3.5 mm, so the test neutron detector could function as a high-efficiency thermal neutron detector as well as a fast neutron spectrometer.  Calculations indicate that approximately 25\% of thermal neutrons incident on the $^{6}$Li-loaded liquid scintillator will backscatter out from the incident surface of the scintillator.  The neutrons that do penetrate into the scintillator volume will undergo neutron capture with nearly 100\% efficiency.

Ambient thermal neutron flux measurements were performed at two locations in the vicinity of the 20 MW research reactor at the NIST Center for Neutron Research (NCNR): outside the concrete shield of the research reactor in the confinement building, and at the end station of the NG-6 cold neutron beamline \citep{Nic05} in the experimental hall.  The beam line extends approximately 70 m from the reactor, and the measurement was taken off-axis of the beam and with the beam shutter closed.  The data acquisition of the test neutron detector was operated without requiring a capture-gated coincidence.  When running in this mode, the detector is also sensitive to gamma-rays so three different shielding configurations were investigated: 1) unshielded; 2) enclosed within a 10-cm thick lead house to suppress the gamma-ray background; and 3) surrounded by 9 mm of borated-silicone to reduce the thermal neutron flux, and enclosed within a 10-cm thick lead house to reduce the gamma-ray background.

\begin{figure*}
    \centering
    $^{6}$Li-loaded Scintillator
    \subfloat{\includegraphics[width=0.5\linewidth]{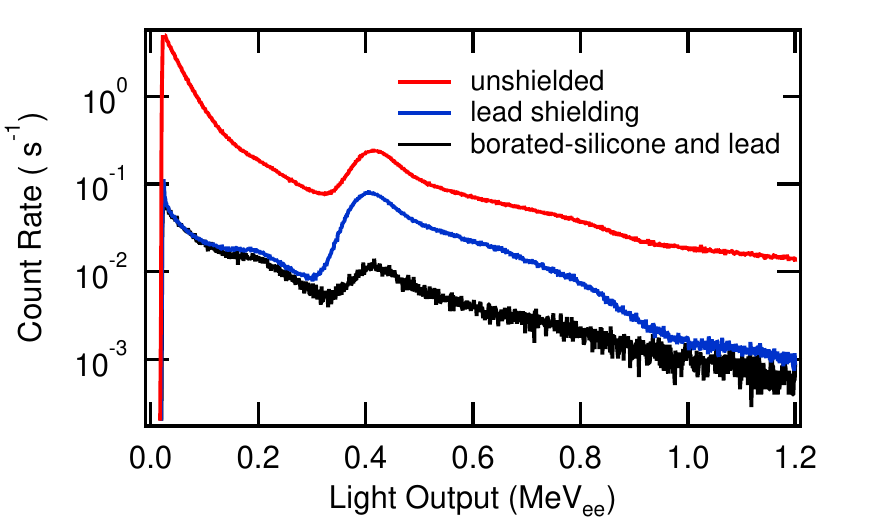}}
    \subfloat{\includegraphics[width=0.5\linewidth]{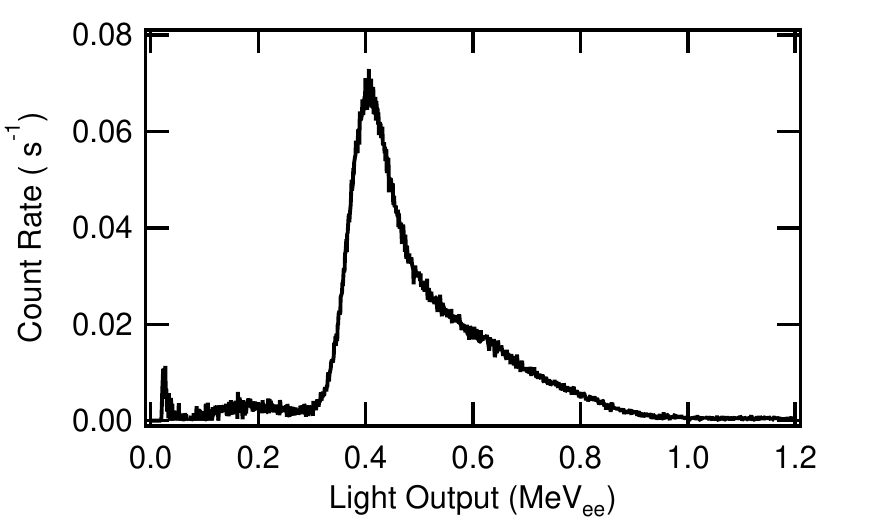}}\\
    Boron-loaded Plastic Scintillator
    \subfloat{\includegraphics[width=0.5\linewidth]{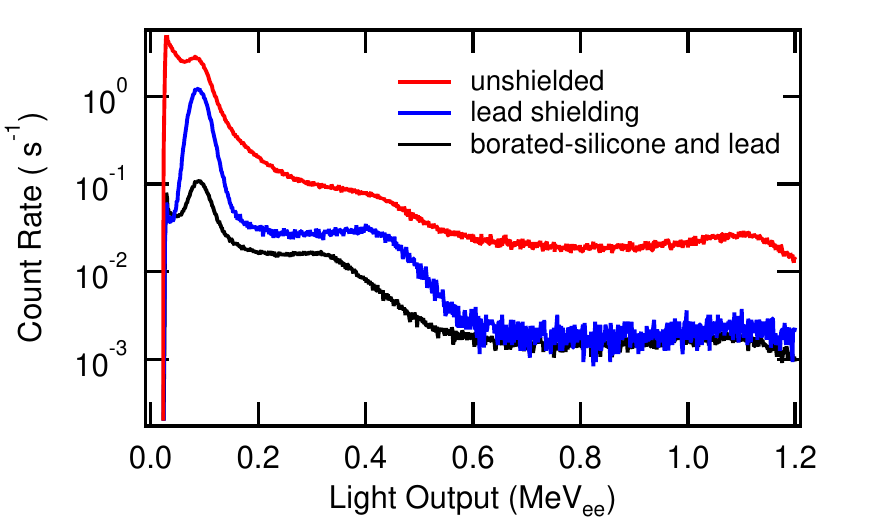}}
    \subfloat{\includegraphics[width=0.5\linewidth]{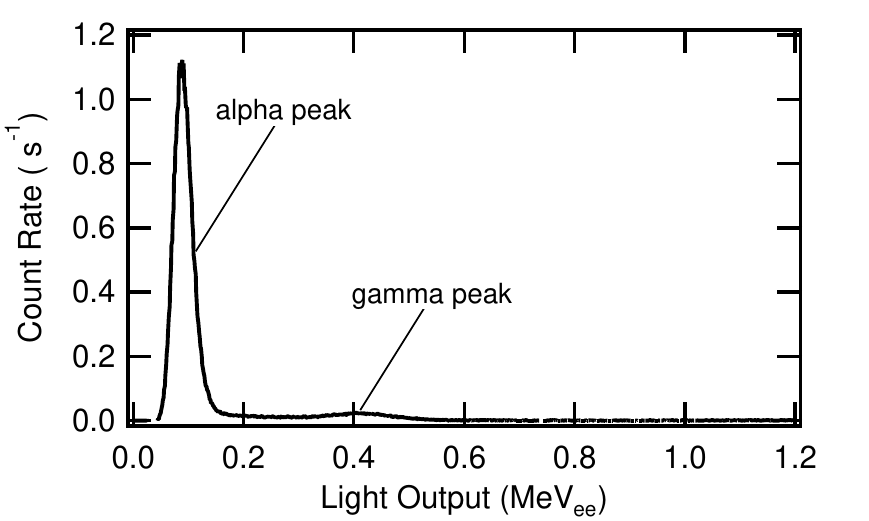}}
    \caption{Ambient thermal neutron flux measured outside the concrete shielding of the research reactor at the NCNR.  The test neutron detectors employed $^{6}$Li-loaded scintillator (top graphs) and BC-454 boron-loaded plastic scintillator (bottom graphs).  The left graphs shows the light output spectra for the different shielding configurations designed to suppress gamma-ray backgrounds and isolate thermal neutrons.  The right graphs show the ambient thermal neutron capture signal spectra after subtracting the borated-silicone and lead shielding spectra from the lead-only shielding spectra.}
\label{C100_thermal}
\end{figure*}

For comparison, an additional measurement at each location and shielding configuration was taken by replacing the $^{6}$Li-loaded scintillator filled glass cell in the test neutron detector with a 5 cm diameter by 6 cm cylindrical block of Saint-Gobain BC-454 natural boron-loaded plastic scintillator.  Calculations indicate the mean free path of thermal neutrons in BC-454 is 2.6 mm and the backscattering of thermal neutrons out of the incident surface is 6\%.  Gamma-ray energy calibrations were performed on each detector using a $^{60}$Co gamma-ray source, and the calibrations were used to convert the pulse height histograms from the thermal neutron flux measurements into light output spectra.  The spectra for the lead and borated-silicone shielding configuration were used as a background subtraction for the lead-only shielding spectra, and the resulting spectra were integrated to yield an ambient thermal neutron flux measurement at each location.  The measured and background-subtracted spectra in the reactor confinement building are shown in Figure~\ref{C100_thermal}.

Because the test neutron detector uses a borosilicate glass cell to contain the $^{6}$Li-loaded liquid scintillator, some fraction of thermal neutrons capture on the naturally-occurring $^{10}$B nuclei in the glass, thus reducing the fluence of thermal neutrons incident on the $^{6}$Li-loaded scintillator.  To quantify the lost fraction, neutron transmission measurements were performed on an empty cell using the Alpha-Gamma neutron detector \citep{Gil89} and the NG-6M neutron beam line at the NCNR \citep{Nic05}.  The NG-6M beam line produces a $\lambda = 0.496$ nm monochromatic neutron beam, and the Alpha-Gamma neutron detector can count the number of neutrons for a sample in-beam and out-beam.  Because the measured neutron fluence can be reduced by both absorption and scattering in the sample, gamma-ray and beta-ray rates were measured with hand-held monitors near the glass cell.  No excessive gamma-ray or beta-ray production was observed, so neutron losses through the glass cell were predominantly due to absorption.

The measured transmission of 0.496 nm neutrons passing through the glass cell was 7.3\%.  However, the $^{6}$Li-loaded scintillator is contained within the interior of the cell so neutrons incident on the scintillator nominally pass through only a single wall thickness of glass, thus the corrected transmission value for the borosilicate glass in the test neutron detector is 27.0\%.  In addition, the neutron capture cross section for $^{10}$B is proportional to $1/v$, so the energy-correction for thermal neutrons ($\lambda = 0.18$ nm) yields a transmission of 44.9\% for the glass cell in the test neutron detector.

After correcting for transmission losses through the borosilicate glass ($^{6}$Li-loaded scintillator only) and back-scattering (approximately 25\% for the $^{6}$Li-loaded scintillator and 6\% for BC-454), both detectors measured an ambient thermal neutron flux of approximately 2.0 $\textrm{cm}^{-2} \textrm{s}^{-1}$ in the confinement building of the research reactor.  A measurement  using a 4-atm $^{3}$He neutron detector-proportional counter measured the ambient thermal neutron flux at this location as 1.9 $\textrm{cm}^{-2} \textrm{s}^{-1}$.  At the end station of the NG-6 cold neutron beamline, both detectors measured an ambient thermal neutron flux of approximately 0.11 $\textrm{cm}^{-2} \textrm{s}^{-1}$.

\section{Discussion}

Because of the microemulsion nature of the loaded scintillator, the $^{6}$Li nuclei are not dispersed uniformly throughout the scintillator bulk but are localized in suspended droplets.  This means that charged particle products from neutron capture on $^{6}$Li -- an alpha and triton -- must first escape from a droplet and then pass through other droplets as they propagate through the scintillator.  The particles deposit energy during their transit through the droplets but do not produce scintillation light.  This loss of light output corresponds to a downward shift in the measured spectra of the neutron capture on $^{6}$Li.  The same process affects the measured proton recoil light output spectra and the gamma-ray calibration spectra.

An estimate of the effect can be calculated based on the volume fraction of aqueous lithium chloride in the $^{6}$Li-loaded scintillator and the ranges of the charged particles in the lithium chloride solution and Quickszint 164.  The ranges for each of the charged particles in 9.40 M lithium chloride solution and Quickszint 164 as calculated using the \emph{Stopping and Range of Ions in Matter} 2008 software package \citep{SRIM} are large compared to the droplet size and are roughly equivalent (see Table~\ref{range}).  An estimate of the loss of light output when compared to unloaded scintillator cocktail is given by the volume fraction of lithium chloride solution, which for the $^{6}$Li-loaded scintillator is 7.5\%.  However, at these energies this reduction is approximately the same for the energy spectra from neutron capture, proton recoil, and gamma-ray sources, so energy calibrations are not significantly affected.

\begin{table}[t]
%\begin{table}[b]
%\begin{table}[htbp]
    \centering
    \begin{tabular}{lcc}
        & 9.40 M LiCl(aq) & Quickszint \\
        \hline
        2.05 MeV alpha & 10.8 $\mu$m & 10.4$\mu$m \\
        2.73 MeV triton & 66.6 $\mu$m & 62.5 $\mu$m \\
        \hline
    \end{tabular}
    \caption{Ranges of the charged particle products of neutron capture on $^{6}$Li through the components of the 0.40\% $^{6}$Li-loaded scintillator as calculated using SRIM.  The calculation assumes a Bragg correction of -6.0\% for the lithium chloride solution and +5.0\% for the liquid scintillator.}
    \label{range}
\end{table}

While the 100 ml test detector demonstrated the ability of $^{6}$Li-loaded scintillator to unambiguously detect capture neutrons and provide incident neutron energy information from the proton recoil spectra, the estimated efficiency was of order $10^{-3}$.  For a fast neutron spectrometer to be of practical value, it would require good efficiency and energy resolution with low sensitivity to uncorrelated background events.  Detector efficiency could be improved by increasing the volume of scintillator because the mean diffusion time for a thermalized neutron to propagate out of the scintillator would increase, which would increase the probability of neutron capture.  In addition, efficiency could be improved by increasing the concentration of $^{6}$Li within the scintillator, which would decrease the mean thermal neutron capture time.

However, increasing the volume of the detector would require optimizing the optical performance of the $^{6}$Li-loaded scintillator to obtain good energy resolution.  The volume fraction of lithium chloride solution that was added to Quickszint 164 was chosen to maximize the mass fraction of $^{6}$Li within the scintillator, maintain microemulsion stability, and still retain good optical properties.  For a spectrometer that incorporates large volumes of scintillator, the choice of volume fraction of lithium chloride should also consider the attenuation length of the scintillator.  In addition, the loss of energy resolution due to the non-linear light yield of organic scintillators could be addressed by employing a multichannel, optically-segmented detector \citep{Abd02}, which would allow better resolution of the energy deposition from multiple elastic scattering during neutron thermalization.

This type of fast-neutron detector should have good efficiency and have energy resolution capable of detecting low-rate signals from fissile materials.  Additional rejection of false events could be accomplished through pulse-shape discrimination techniques \citep{Fla07,Fis11,Kle02,Wol95} and using energy information in analysis.  It should be noted that acquisition time needed to produce the fast neutron energy spectra detailed in this paper was of order 10 hours and was a consequence of the small volume of the $^{6}$Li-loaded scintillator.  Any practical device for detecting low-rate signals from fissile materials would require a spectrometer that incorporates several liters of $^{6}$Li-loaded scintillator.  MCNP5 studies indicate that a fast neutron spectrometer containing about 10 liters of $^{6}$Li-loaded scintillator in a cubic geometry could detect neutrons from unmoderated $^{252}$Cf at a rate of 0.3 $\textrm{ng}^{-1}\textrm{s}^{-1}$ at a distance of 2 m in the energy range of 1 MeV to 20 MeV.

As a thermal neutron detector, the $^{6}$Li-loaded scintillator performed comparable to commercially-available boron-loaded scintillator, and both scintillators yielded ambient thermal neutron flux measurements that agree at the 10\% level.  However, the neutron capture signal for the $^{6}$Li-loaded scintillator was well-separated from the low-energy gamma-ray tail in the unshielded detector configuration, while the neutron capture signal (the primary alpha peak) for the BC-454 boron-loaded plastic was well within the low-energy gamma-ray tail in its unshielded detector configuration (see Figure~\ref{C100_thermal}).  This separation of the neutron capture by the $^{6}$Li-loaded scintillator could allow an analysis of the gamma-ray spectra concurrently with the thermal neutron flux measurement, which might not be possible with boron-loaded scintillators (and is not possible with a $^{3}$He neutron detector-proportional counter).

\section{Conclusion}

We have developed a $^{6}$Li-loaded liquid scintillator suitable for use in fast neutron spectrometry and thermal neutron detection. We chose the $^{6}$Li-loading based on optical transmittance within a wavelength band appropriate for PMT operation and detector efficiency. Irradiations of a test spectrometer by 2.5 MeV neutrons, 14.1 MeV neutrons, and neutrons from $^{252}$Cf decay demonstrated that the scintillator is capable of cleanly identifying neutron capture events and generating a proton recoil light output spectrum that is related to the incident neutron energy. In addition, the $^{6}$Li-loaded scintillator was used to measure ambient thermal neutron flux and performed comparably to a boron-loaded plastic scintillator.

A $^{6}$Li-loaded liquid scintillator has some advantages over scintillators loaded with other neutron capture isotopes and may be the preferred agent for some applications.  The products from the $^{6}$Li$(n,\alpha)^{3}$H reaction are charged particles and not gamma-rays, so their energy deposits are completely contained within the scintillating medium.  This is particularly advantageous if the detection volume is not large.  The \emph{Q}-value of the $^{6}$Li$(n,\alpha)^{3}$H reaction is larger than that for $^{10}$B$(n,\alpha)^{7}$Li, so the energy deposit peak is much better separated from the noise threshold.  The production of aqueous lithium chloride uses straightforward chemical procedures and does not require elaborate facilities.  In addition, the cost of loading the liquid scintillator with $^{6}$Li is significantly less than commercially available boron-loaded scintillators.  We do note that acquiring enriched $^{6}$Li may prove difficult for some institutions.

The present investigations show promise for $^{6}$Li-loaded scintillator, but further research into its properties is warranted.  As a thermal neutron detector, one should quantify the effect of gamma-ray contamination in the capture peak; pulse shape discrimination methods could optimize the neutron-to-gamma signal in that region.  The optical and physical quality of the loaded-scintillator has been observed over the course of approximately one-year, but for most applications, it is reasonable to think that researchers would want stability over periods of several years.  In addition, these investigations have used only relatively small samples; for a large-scale detector, one must know the attenuation length of the loaded scintillator and its stability over time.

\section{Acknowledgements}

The authors acknowledge the support of the National Institute of Standards and Technology, U.S. Department of Commerce, in providing the neutron research and chemistry facilities used in this work.  This work is supported in part by NSF PHY-0809696 and NSF PHY-0757690.  Tom Langford acknowledges support under the National Institute of Standards and Technology American Recovery and Reinvestment Act Measurement Science and Engineering Fellowship Program Award 70NANB10H026 through the University of Maryland.

We acknowledge Dr. Andrew Yue at the NIST Center of Neutron Research for his help with cold neutron transmission studies.  We acknowledge Dr. Paul DeRose at the NIST Advanced Chemical Sciences Laboratory for his help with fluoroscopic measurements.  We thank Dr. Vladimir Gavrin and Dr. Johnrid Abdurashitov of the Institute for Nuclear Research - Russian Academy of Sciences for useful discussions.

\end{document}